\documentstyle[aps,psfig,amsmath]{revtex}
\newcommand{\be}{\begin{equation}}
\newcommand{\ee}{\end{equation}}
\newcommand{\bea}{\begin{eqnarray}}
\newcommand{\eea}{\end{eqnarray}}
\newcommand{\bef}{\begin{figure}}
\newcommand{\eef}{\end{figure}}
\newcommand{\bm}{\bibitem}
\newcommand{\al}{\alpha}
\newcommand{\bet}{\beta}
\newcommand{\gm}{\gamma}
\newcommand{\de}{\delta}
\newcommand{\lm}{\lambda}
\newcommand{\sg}{\sigma}
\newcommand{\gf}{\gamma_5}
\newcommand{\ep}{\epsilon}

\newcommand{\om}{\omega}
\newcommand{\rw}{\rightarrow}
\newcommand{\Rw}{\longrightarrow}

\newcommand{\cl}{{\cal{L}}}
\newcommand{\cj}{{\cal{J}}}
\newcommand{\F}{F_\pi}
\newcommand{\Fr}{F_\rho}
\newcommand{\dmd}{\partial_\mu}

\newcommand{\dnu}{\partial^\nu}

\newcommand{\ps}{p \!\!\! /}
\newcommand{\ks}{k \!\!\! /}
\newcommand{\us}{u \!\!\! /}

\newcommand{\bu}{\bar{u}}

\newcommand{\la}{\langle}
\newcommand{\ra}{\rangle}
\renewcommand{\d}{\cdot}

\newcommand{\pr}{^\prime}

\begin{document}

\setcounter{page}{1}

\title{Current algebra derivation of temperature dependence of hadron couplings with currents}

\author{S. Mallik} 
\address{Saha Institute of Nuclear Physics,
1/AF, Bidhannagar, Kolkata-700064, India} 

%\date{} 

\maketitle

\begin{abstract} 

The vector and the axial-vector meson couplings with the vector and the
axial-vector currents respectively at finite temperature have been
obtained in Ref. \cite{Mallik} by calculating all the relevant one-loop 
Feynman graphs with vertices obtained from the effective chiral Lagrangian. 
On the other hand, the same couplings were also derived in Ref.\cite{Ioffe1} 
by applying the method of current algebra and the hypothesis of partial 
conservation of axial-vector current (PCAC). The latter method appears to 
miss certain terms; in the case of the vector meson coupling with the vector 
current, for example, a term containing the $\rho\omega\pi$ coupling is 
missed. A similar situation would also appear for the nucleon coupling with 
the nucleon current. In this note we resolve these differences. 

\end{abstract}

\section{Introduction}
\setcounter{equation}{0}
\renewcommand{\theequation}{1.\arabic{equation}}

The thermal two-point functions of the vector and the  axial-vector 
currents are evaluated in Ref. \cite{Mallik} from their 
one-loop Feynman graphs with vertices given by  the effective chiral
Lagrangian of strong interactions \cite{Ecker}. At low, but not too small, 
temperature $T$, the $\rho$ meson coupling  $\Fr$ with the vector current, 
for example, is found to depend on temperature as
\be
\Fr^{T} = \Fr \left \{ 1-\left (1+\frac{g_1^2}{3}\right )
\frac{T^2}{12\F^2} \right \}\, . 
\ee
Here $\F$ is the familiar pion decay constant, $\F =93 MeV$ and $g_1$ is
the $\rho\om\pi$ coupling, $ g_1 = .87$. Although there is no strict power
counting of momenta according to the number of loops when heavy mesons are
present, the above result is expected to hold to within an accuracy of 
about 20\% \cite{Ecker}.

Earlier, the authors of Ref. \cite{Ioffe1} also evaluated the same 
quantities in a somewhat different way. At low temperature they expanded
the thermal trace over states in the two-point functions, retaining only 
the vacuum and the one pion state. The forward pion-current amplitude so 
obtained was apparently evaluated with the methods of current algebra and 
PCAC, reducing it to a combination of vacuum two-point functions of 
the vector and the axial-vector currents. It allowed them to conclude
that the coupling $\Fr^T$ would change with temperature according to 
Eq.(1.1), but {\it without} the $g_1^2$ term.
 
A similar situation would arise also in the calculation of the temperature
 dependence of the nucleon coupling $\lm$ with the nucleon current 
\cite{Ioffe2}. It has been calculated in Ref.\cite{Leutwyler1} as
\be
\lm^{T} = \lm \left \{ 1-\frac{(g_A^2 +1)}{32} \frac{T^2}{\F^2} \right\}\,,
\ee
by evaluating in the effective theory the one-loop Feynman graphs for
the thermal two-point function of nucleon currents. Here $g_A$ is the 
axial-vector coupling constant of the nucleon, $g_A=1.26$. If, however, 
we follow  the procedure of Ref. \cite{Ioffe1} to work out the same 
quantity by applying apparently the current algebra techniques, we would 
{\it miss} the $g_A^2$ term in Eq.(1.2). 

The evaluation of the thermal part of one-loop Feynman graphs is
actually a tree level calculation, as the corresponding part of the
propagator of a particle (here pion) contains a mass-shell delta function.  
Now, at the tree lavel, the current algebra plus PCAC approach is strictly 
equivalent to the method of chiral perturbation theory \cite{Leutwyler2}. 
One thus wonders why the two approaches would lead to different results for 
the temperature dependence of the couplings stated above.

Historically, the idea of PCAC was introduced  \cite{Gell-Mann} to extract
the pion pole contribution to matrix elements of operators containing the
axial-vector current. At small momentum $k$ carried by the current, this is
the only source of dominant ciontribution, as far as processes involving
only pions (Goldstone bosons) are concerned. However, when there are heavy
(non-Goldstone) particles in the process, there is another potential source
of dominant contribution to such matrix elements, namely the graphs, in
which the axial-vector current attaches to an external heavy particle line. 
The additional heavy particle propagator so created can be singular at small 
momentum $k$. In fact, this is the mechanism contributing dominantly to 
processes of soft pion emission and absorption \cite{Nambu}.

In this note we rederive the temperature dependence of $\Fr^T$ and $\lm^T$
in the framework of current algebra in the chiral symmetry limit. We begin 
with the Ward identities for 
the relevant amplitudes involving the pion and the currents. We show that 
there are indeed heavy particle poles in the above two-point functions 
that are as dominant as the pion pole at small momentum $k$. In the case of
the vector current, it is the $\om$ meson pole in the neighbourhood of
the $\rho$ meson pole, when we consider temperatures higher than the 
mass difference of the two mesons. In the case of nucleon current, it is
the nucleon pole itself, there being no mass difference here
to deal with. It is these contributions, giving rise to the $g_1^2$ and the
$g_A^2$ terms in Eqs. (1.1-2) respectively, that are ignored in 
Ref. \cite{Ioffe1}. 

In Secs. II and III we present current algebra derivations of Eqs. (1.1)
and (1.2) respectively, where the contributions of the heavy particle poles
are also calculated explicitly along with that of the pion pole. Our discussions 
are contained in Sec.IV.

\section{Current Algebra derivation of Eq.(1.1)} 
\setcounter{equation}{0}
\renewcommand{\theequation}{2.\arabic{equation}}

The derivation is based on the thermal two-point function of vector
currents, $V_\mu^i(x), i=1,2,3$, with which the $\rho$ communicates. At
low temperature $\bet^{-1}(=T)$, the heat bath is dominated by pions. Thus
the leading correction to the $\rho$ meson propagation is expected to come
from its scattering off the pions. Accordingly the thermal trace is expanded
as \cite{Ioffe1},
\bea
& &Tre^{-\bet H} TV_\mu^i(x)V_\nu^j(0)/Tre^{-\bet H} \nonumber \\  
& & =\la 0|TV_\mu^i(x)V_\nu^j(0)|0\ra + \int\frac {d^3k\, n(k)}{(2\pi)^3\,
2|\vec{k}|} 
\sum_{m} \la\pi^m(k)|TV_\mu^i(x)V_\nu^j(0)|\pi^m(k)\ra_{conn},
\eea
where $n(k)=(e^{\bet |\vec{k}|} -1)^{-1} $ is the distribution function of
pions. $T$ also denotes the time ordering of all operators that follow
it. $\la \cdots \ra_{conn}$ is the connected part of the matrix element.
The first term gives the $\rho$ pole in terms of its mass and coupling in 
vacuum. To find their changes at finite temperature to leading order,
we use the method of current algebra to extract the leading terms of the 
pion matrix element at small $k$.

\bef
\centerline{\psfig{figure=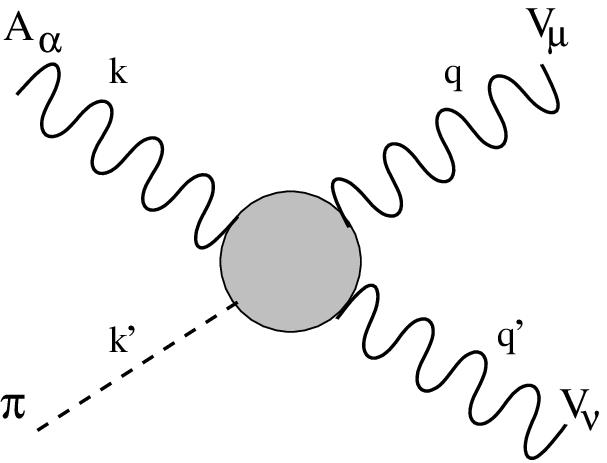,height=3.0cm,width=4.0cm}}
\caption{Amplitude $W^{mnij}_{\al\mu\nu}(q,k,q',k')$ describing the
scattering $\pi (k') + V_\nu(q')\Rw A_\al (k)+V_\mu (q)$. Isospin indices
are omitted in the diagram.}
\eef

The appropriate Ward identity follows from the momentum space amplitude
\be
W^{mnij}_{\al\mu\nu} (q,k,q',k')= \int d^4z d^4x d^4x' e^{ik\d z} e^{iq\d x} 
e^{-iq'\d x'}\la 0|TA^m_\al (z)V_\mu^i(x)V_\nu^j(x')|\pi^n(k') \ra\, ,
\ee
describing the scattering process shown in Fig.1. As usual, we contract it
with $k^\al$ and integrate by parts with respect to $z$ and
ignore the surface term to get
\be
k^{\al} W^{mnij}_{\al\mu\nu} (q,k,q',k')= i\int d^4zd^4xd^4x'e^{ik\d z}
e^{iq\d x} e^{-iq'\d x'}
\partial^{\al}_z  \la 0|TA^m_\al (z)V_\mu^i(x)V_\nu^j(x')|\pi^n(k') \ra\,.
\ee
Since the axial-vector current is conserved in the chiral limit, the divergence 
operator gives non-zero contributions only when it acts on the theta functions 
in the definition of the time-ordered product, giving
\begin{eqnarray*}
& & \partial^{\al}_z  \la 0|TA^m_\al (z)V_\mu^i(x)V_\nu^j(x')|\pi^n(k')\ra
\nonumber \\
& &=\de (z_0-x_0)\la 0|T([A_0^m(z),V_\mu^i(x)]V_\nu^j(x'))|\pi^n(k')\ra
+\de (z_0-x'_0)\la 0|T([A_0^m(z),V_\nu^j(x')]V_\mu^i(x))|\pi^n(k')\ra . 
\end{eqnarray*}
The equal time commutators of currents are given by current algebra,
\[[A^m_0(z),V_\mu^i(x)]_{z_0=x_0} =i\ep^{mil} A^l_\mu (x) \de^3(z-x)\, . \]
Also using translational invariance to remove one more integration in Eq.
(2.3), we get the Ward identity as
\bea
& & k^{\al} W^{mnij}_{\al\mu\nu} =-(2\pi)^4 \de^4 (q+k-q'-k') \nonumber  \\
& & [\ep^{mil}\int d^4x e^{i(q+k)\d x} \la 0|TA_\mu^l(x)V_\nu^j(0)|\pi^n(k')\ra
+\ep^{mjl}\int d^4xe^{iq\d x}\la 0|TV_\mu^i(x)A_\nu^l(0)|\pi^n(k')\ra ]\,.
\eea

To find the leading (constant) piece in $k^\al W^{mnij}_{\al\mu\nu}$ in the
limit $k^\al \rw 0$, we need only to calculate the terms in 
$W^{mnij}_{\al\mu\nu}$ that are singular in this limit. One such term is, 
of course, the $\pi$ pole contribution, as shown in graph (a) of Fig. 2, 
that can be readily evaluated by inserting the pion intermediate state in
the matrix element of Eq.(2.2),
\be
k^{\al} W^{mnij}_{\al\mu\nu} \stackrel{\pi}{\Rw} 
- (2\pi)^4\de^4(q+k-q'-k')\F \int d^4x e^{iq\d x} \la
\pi^m(k)|TV_\mu^i(x)V_\nu^j(0)|\pi^n(k')\ra ,
\ee
where $\F$ is defined by 
\[\la 0|A_\al^m(z)|\pi^n(k) \ra =i\de^{mn}\F k_\al e^{-ikz}. \]
As we shall show below, there is another finite contribution to $k^{\al}
W^{mnij}_{\al\mu\nu}$ as $k^{\al} \rw 0$ due to the $\om$ pole in the vicinity
of the $\rho$ pole. Denoting this contribution as
\[k^{\al} W^{mnij}_{\al\mu\nu} \stackrel{\om}{\Rw}
(2\pi)^4\de^4(q+k-q'-k') k^{\al} R_{\al\mu\nu}^{mnij}(q,k,k'), \]
the Ward Identity (2.4) gives the result
\bea
& &\F\int d^4x e^{iq\d x} \la  \pi^m(k)|TV_\mu^i(x)V_\nu^j(0)|\pi^n(k')\ra 
\nonumber \\
& &=\ep^{mil}\int d^4x e^{i(q+k)\d x} \la 0|TA_\mu^l(x)V_\nu^j(0)|\pi^n(k')\ra 
+\ep^{mjl}\int d^4x e^{iq\d x} \la 0|TV_\mu^i(x)A_\nu^l(0)|\pi^n(k')\ra
+k^\al R_{\al\mu\nu}^{mnij}\, .
\eea

The first two matrix elements on the right of Eq.(2.6) can be reduced to vacuum 
matrix elements, again by using Ward identities. Thus for the first one
we define the amplitude
\be
w_{\bet\mu\nu}^{mnij} (q,k,k')=\int d^4z'd^4x e^{-ik\pr\d  z\pr}e^{i(k+q)\d x} 
\la 0|TA_\mu^l(x)V_\nu^j(0)A_\bet^n(z')|0\ra
\ee
One proceeds exactly as before by contracting $w_{\bet\mu\nu}^{mnij}$ with
$k'^\bet$. Here the pion pole is the only dominant contribution at small
$k'$, so that the Ward identity for this amplitude becomes
\bea
& &\F\int d^4x e^{i(k+q)\d x} \la 0 |TA_\mu^l(x)V_\nu^j(0)|\pi^n(k')\ra 
\nonumber \\
& &=\ep^{nll\pr}\int d^4x e^{iq\pr\d x} \la 0|TV_\mu^{l\pr} (x)V_\nu^j(0)|0\ra
+\ep^{njl\pr}\int d^4x e^{i(q+k)\d x} \la 0|TA_\mu^l(x)A_\nu^{l\pr} (0)|0\ra \,.
\eea 
A similar equation holds for the second matrix element on the right of
Eq.(2.6). We thus get from Eq.(2.6) the pion matrix element in Eq.(2.1) as 
\bea
& &\int d^4x e^{iqx} \sum_m \la \pi^m(k)|TV_\mu^i(x)V_\nu^j(0)|\pi^m(k)\ra
\nonumber \\
& &=-\frac{4}{\F^2}\int d^4x e^{iq\d x} \la 0|TV_\mu^i(x)V_\nu^j(0)|0\ra
+\frac{4}{\F^2}\int d^4x e^{iq\d x} \la 0|TA_\mu^i(x)A_\nu^j(0)|0\ra
+\frac{1}{\F}k^\al R_{\al\mu\nu}^{ij}(q,k)\,.
\eea

\bef
\centerline{\psfig{figure=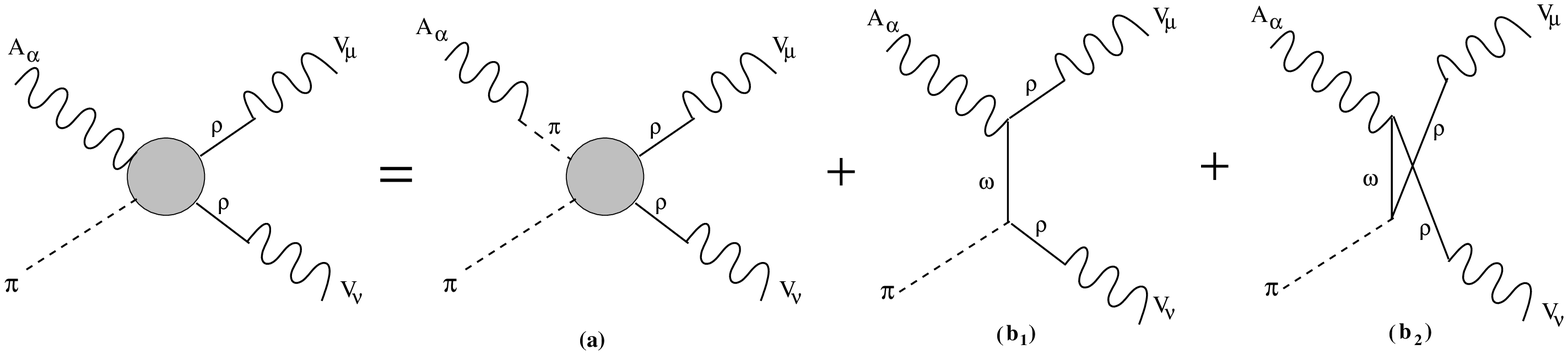,height=3.6cm,width=15.5cm}}
\caption{Singular contributions to $ W^{mnij}_{\al\mu\nu}(q,k,q',k')$
as $k,\,k'\rw 0$, when the vector currents are coupled to $\rho$ meson.} 
\eef

The remaining task is to evaluate the $\om$ meson pole contribution, 
shown in graphs $(b_1)$ and $(b_2)$ of Fig. 2. However, we do not evaluate 
them as Feynman graphs with the intermediate lines as propagators, but insert
intermediate states in the spirit of the method of current algebra, as we 
already did with the pion pole extraction in
Eq.(2.5). Noting the quantum numbers of $\om$, it is given by
\bea
& & W^{mnij}_{\al\mu\nu}(q,k,q',k') \stackrel{\om}{\Rw}\int d^4z d^4x d^4x'
e^{ikz}e^{iqx}e^{-iq'x'} \nonumber \\ 
& & \sum_{\sg_1}\int \frac{d^4q_1}{(2\pi)^4} 
\la 0|TA_\al^m(z)V_\mu^i(x)|\om(q_1,\sg_1)\ra
\la\om(q_1,\sg_1)|V_\nu^j(x')|\pi^n(k')\ra \frac{i}{q_1^2-m_\om^2}
+\mathrm{crossed \,\,  term}.
\eea
Its correction to the $\rho$ pole is obtained by isolating this pole
at the vector current vertices. Then extracting the coordinate dependence
by translation invariance, we can integrate over the coordinates to get
delta functions in momentum, which are then removed by intergals over
the intermediate momenta. We thus get
\bea
& & W_{\al\mu\nu}^{mnij} \stackrel{\om,\rho}{\Rw} 
\sum_{\sg_1,\sg_2,\sg_3,b,c}\la 0|V_\mu^i(0)|\rho^b(q,\sg_2)\ra 
\la\rho^b(q,\sg_2)|A_\al^m(0)|\om(q+k,\sg_1)\ra \nonumber \\
& & \la\om(q+k,\sg_1)|\pi^n(k'),\rho^c(q',\sg_3)\ra \la
\rho^c(q',\sg_3)|V_\nu^j(0)|\ra\frac{i^3}{(q^2-m_\rho^2)(q'^2-m_\rho^2)
\{(q+k)^2-m_\om^2\}} +\mathrm{crossed \,\, term}\,.
\eea

To evaluate the matrix elements in Eq.(2.11), we write the relevant pieces 
of the effective Lagrangian based on the chiral symmetry of QCD. In the 
notation of Ref.\cite{Ecker}, these are
\be
\cl_{int} = \frac{1}{2\sqrt{2}}\frac{F_V}{m_V} 
\la \hat{V}_{\mu\nu}f_+^{\mu\nu}\ra +\frac{\sqrt{3}}{4}
g_1\ep_{\mu\nu\rho\sg}\la\hat{V}^{\mu\nu}\{\hat{V}^\rho,u^\sg\}\ra.
\ee
Here the fields are all $3\times 3$ octet matrices. $\hat{V}^\mu$ denotes 
the vector meson fields and $\hat{V}^{\mu\nu}$ their field strengths. 
The field strengths
$f_+^{\mu\nu}$ (and $f_-^{\mu\nu}$) are those of external fields, $v_\mu^i$
and $a_\mu^i$, with which the vector and the axial-vector currents are
coupled in the QCD Lagrangian. $u^\mu$ is related to the  covariant derivative 
of the pseudoscalar (Goldstone) fields. All these matrix valued fields transform 
only under the unbroken subgroup of QCD. Taking the trace denoted by 
$\la\cdots\ra$, the above terms are actually invariant under the full chiral 
symmetry group. Restricting 
to terms needed in our calculation, Eq.(2.12) gives immediately the vector 
and the axial-vector currents in the effective theory   
$[u\overleftrightarrow\dmd v\equiv (\dmd u) v-u\dmd v]$,  
\[V_\mu^i=F_\rho m_\rho \rho_\mu^i~,~~~~~~ A_\sg^i =- g_1\ep_{\mu\nu\lm\sg}
\om^\mu\overleftrightarrow\dnu \rho^{\lm i}\, , \]
and also the $\rho\om\pi$ interaction, 
\[\cl_{\rho\om\pi}= \frac{g_1}{\F}\ep_{\mu\nu\lm\sg}
 \om^\mu\overleftrightarrow\dnu \rho^{\lm i}\partial^\sg \pi^i\,, \]
where the symbols denote the corresponding fields.
Then we have for the matrix elements 
\begin{eqnarray*}
& & \la0|V_\mu^i(0)|\rho^b(q,\sg_2)\ra=\de^{ib} F_\rho m_\rho
\ep_\mu(q,\sg_2)\,, \\
& & \la\rho^b(q,\sg_2)|A_\al^m(0)|\om(q+k,\sg_1)\ra =i\de^{mb}g_1
\ep_{\mu\nu\lm\al}(2q+k)^\nu \ep^{\lm *}(q,\sg_2)\ep^\mu(q+k,\sg_1)\,,
\end{eqnarray*}
where $\ep^\mu(q,\sg)$ is the polarisation state of the vector meson with
momentum $q$ and spin projection $\sg=(1,0,-1)$. The remaining matrix
element is a S-matrix element, which to first order in perturbation
expansion is,
\begin{eqnarray*}
& & \la\om(q+k,\sg_1)|\pi^n(k'),\rho^c(q',\sg_3)\ra \\
& & =(2\pi)^4\de (q+k-q'-k') \de^{nc}i\frac{g_1}{\F} \ep_{\mu\nu\lm\sg}
(2q+2k-k')^\nu k^{\prime \sg} \ep^\lm (q',\sg_3)\ep^{\mu *}(q+k,\sg_1)\,.
\end{eqnarray*}

Inserting these matrix elements in Eq.(2.11), we can carry out the sums 
over the polarisation states of the vector mesons. After a little
simplification, we get
\be
\frac{1}{\F}k^\al R_{\al\mu\nu}^{ij} = -4i\de^{ij}\left(\frac{g_1}{\F}\right)^2
\frac{\Lambda_{\mu\nu}}{(q^2-m_\rho^2)^2} \left \{\frac{1}{(q+k)^2-m_\om^2}
+\frac{1}{(q-k)^2-m_\om^2} \right\}\,,
\ee
where the Lorentz tensor $\Lambda_{\mu\nu}$ is given by
\begin{eqnarray*}
\Lambda_{\mu\nu}(q,k) &=&\ep_{\al\bet\gm\mu}\ep^\al_{\bet\pr\gm\pr\nu} q^\bet
q^{\bet\pr}k^\gm k^{\gm\pr}\\
&=& (q.k)^2g_{\mu\nu} +q^2k_\mu k_\nu -(q.k)(k_\mu q_\nu + k_\nu q_\mu)\,,
\end{eqnarray*}
in the chiral limit ($k^2=0$). We can now write the complete result of our 
calculation as
\bea
& i &\int d^4x e^{iq\d x}Tre^{-\bet H} TV_\mu^i(x)V_\nu^j(0)/Tre^{-\bet H} \nonumber  \\
& = &\left( 1-\frac{4\cj}{\F^2} \right) i\int d^4x e^{iq\d x} 
\la 0|TV_\mu^i(x)V_\nu^j(0)|0\ra +
\frac{4\cj}{\F^2} \int d^4x e^{iq\d x} \la 0|TA_\mu^i(x)A_\nu^j(0)|0\ra \nonumber \\
& + & 8\de^{ij} \left(\frac{g_1}{\F}\right)^2
\frac{(q^2-m_\om^2)}{(q^2-m_\rho^2 +i\ep)^2}\int\frac{d^3k\, n(k)}{(2\pi)^3\,
2|\vec{k}|} \frac{\Lambda_{\mu\nu }(q,k)}{(q^2-m_\om^2+i\ep)^2 -4(q\d k)^2},
\eea
where we reinstate $i\ep$ to define the poles and
\[\cj=\int\frac{d^3k\, n(k)}{(2\pi)^3\,2|\vec{k}|} = \frac{T^2}{24}\,.\]

To proceed with the evaluation, we have to take note of the 
kinematics of the thermal two-point functions. Any such function can be 
decomposed into two invariant amplitudes. We now set $\vec{q}=0$ with 
$q_0\equiv E$, when the two amplitudes as functions of $E$ become related 
to each other. Thus it suffices for us to consider, for example, 
the trace of the two-point function in its Lorentz indices. The free $\rho$ 
pole term in vacuum then reads as
\be 
i\int d^4x e^{iqx} \la 0|TV_\mu^i(x)V^{\mu j}(0)|0\ra \stackrel{\rho}{\Rw}
\frac{3\de^{ij}(\Fr m_\rho)^2}{E^2 -m_\rho^2}\,.
\ee
Next we find this trace for the third term on the right in Eq.(2.14) in the
vicinity of the $\rho$ pole. If we ignore the mass difference between $\rho$
and $\om$ mesons, it can be immediately evaluated to reproduce the $g_1^2$
term in Eq.(1.1). In the realistic case with $\Delta m = m_\om -m_\rho =12\,$ 
MeV \cite{Comment}, it is useful to note that the $k$-integral is
essentially cut off at $|\vec{k}|=T$, because of the pion distribution 
function present in it. Thus if we are interested in the behaviour of $F_\rho^T$ 
for temperatures larger than $ \Delta m$, as is usually the case, we
can again ignore $\Delta m$ to leading order to get Eq.(1.1). But for
temperatures small compared to $\Delta m$, the $k$-integral is $ O(T^4)$ and the 
$g_1^2$ term will not appear in Eq.(1.1).

\section{Current Algebra derivation of Eq.(1.2)}
\setcounter{equation}{0}
\renewcommand{\theequation}{3.\arabic{equation}}

The properties of the nucleon can be studied by constructing a nucleon
current $\eta^a_A (x)$ out of the quark fields, having the quantum 
numbers of the nucleon \cite{Ioffe2}. Here $a,b,\cdots$  and $A, B,\cdots$            
denote respectively the isospin and the Dirac spinor indices. To study its
properties at low temperature, we again expand, as in the case of vector
currents, the thermal two-point function of nucleon currents,
\be
Tre^{-\bet H}T\eta^a_A(x)\bar{\eta}^b_B(0)/Tre^{-\bet H}\,,
\ee
in terms of the vacuum and the one pion state, and apply the method of
current algebra to estimate the resulting pion matrix element of these
currents,
\be
\sum_m \la\pi^m(k) |T\eta^a_A(x) \bar{\eta}^b_B (0) |\pi^m(k) \ra\, ,
\ee
for small pion momenta in the chiral limit. As expected the calculation is
similar to the earlier one with vector currents, except for the fermionic
nature of the nucleon current.

The required Ward identity follows from the amplitude
\be
Z^{mnab}_{\al, AB} (p,k,p',k')= \int d^4z d^4x d^4x' e^{ik\d z} e^{ip\d x}
e^{-ip\d x'} \la 0|TA^m_\al(z)\eta^a_A(x)\bar{\eta}^b_B(0) |\pi^n(k')\ra
\ee
Again using the axial-vector current conservation and the equal time 
commutation relation,
\[ [A^m_0(z),\eta^a_A(x)]_{z_0=x_0} = -\left(\frac{\tau^m}{2}\right)^{ac}
(\gf)_{AC}\eta^c_C(x) \de^3(z-x) \, ,\]
we follow the steps as before to get the Ward identity,
\bea
& & k^\al Z^{mnab}_{\al, AB} =-i(2\pi)^4 \de^4(p+k-p'-k') \nonumber \\
& & \left[ \left(\frac{\tau^m}{2}\right)^{ac}(\gf)_{AC}\int d^4x e^{i(p+k)\d x} 
\la 0 |T\eta^c_C(x)\bar{\eta}^b_B(0) |\pi^n(k')\ra +\int d^4x e^{ip\d x}
\la 0 |T\eta^a_A(x)\bar{\eta}^c_C(0)|\pi^n(k')\ra
\left(\frac{\tau^m}{2}\right)^{cb}(\gf)_{CB} \right]
\eea
The dominant contribution to $Z^{mnab}_{\al, AB}$ at small $k,k'$ arise from
the pion and the nucleon poles. Denote the latter contribution as
\[Z^{mnab}_{\al,AB} \stackrel{N}{\Rw} (2\pi)^4 \de^4(p+k-p'-k')
S^{mnab}_{\al,AB}(p,k,p')\, . \]
The forward scattering amplitude (3.2) can now be obtained as
\bea
& & \int d^4x e^{ip\d x}\sum_m \la\pi^m(k) |T\eta^a_A(x) \bar{\eta}^b_B
(0)|\pi^m(k)\ra \nonumber \\
& & = -\frac{3}{2\F^2}\int d^4x e^{ip\d x}\la 0|T\eta^a_A(x)\bar{\eta}^b_B |0\ra
 -\frac{3}{2\F^2}\int d^4x e^{ip\d x}\la
0|T\tilde{\eta}^a_A(x)\bar{\tilde{\eta}}^b_B |0\ra +\frac{1}{\F} k^\al
S^{ab}_{\al,AB}(p,k)\,,
\eea
where $\tilde{\eta}=\gf \eta$.

It remains to evaluate the contribution of the nucleon intermediate state to
$Z^{mnab}_{\al, AB}$. Of course, it must be inserted twice more to couple it 
to the nucleon currents. We get
\bea
& & Z^{mnab}_{\al, AB}\stackrel{N}{\Rw}\sum_{\sg_1,\sg_2,\sg_3,c,d,e}
\la 0|\eta^a_A(0)|N^d(p,\sg_2)\ra \la N^d(p,\sg_2)|A^m_\al
(0)|N^c(p+k,\sg_1)\ra \,. \nonumber \\ 
& & \la N^c(p+k,\sg_1)|N^e(p',\sg_3),\pi^n(k')\ra
\la N^e(p',\sg_3)|\bar{\eta}^b_B(0)|0\ra\frac{i^3}{(p^2-m_N^2)(p^{\pr 2}
-m_N^2)\{(p+k)^2-m_N^2\}} +\mathrm{crossed\,\, term}
\eea
The coupling of $\eta$ to nucleon is given by the matrix element
\[ \la 0|\eta^a_A(0)|N^d(p,\sg)\ra =\de^{ad} \lm u_A(p,\sg)\,, \]
where $u(p,\sg)$ is the Dirac spinor of the nucleon.
The piece in the effective Lagrangian
\[\cl_{int}=\frac{1}{2} g_A \bar{\psi} \us\gf \psi\]
gives the nucleon contribution to the axial-vector current and the $\pi N
N$ coupling. We then have
\begin{eqnarray*}
& & \la N^d(p,\sg_2) |A^m_\al (0)|N^c(p+k,\sg_1)\ra =\frac{1}{2} g_A
\bu(p,\sg_2)\gm_\al \gf u(p+k,\sg_1)\chi^{d \dag}\tau^m\chi^c \, \\
& & \la N^c(p+k,\sg_1) |N^e(p',\sg_3), \pi^n(k')\ra =(2\pi)^4 \de^4
(p+k-p'-k')\left( -\frac{g_A}{2\F}\right) \bu(p+k,\sg_1)\ks\pr\gf u(p',\sg_3)
\chi^{c \dag}\tau^n\chi^e\,,
\end{eqnarray*}
where $\chi$'s denote isospin states of the nucleon.
Inserting these matrix elaments in Eq.(3.6) and carrying out the spin and 
isospin summations, we get
\bea
S^{mnab}_{\al, AB} = & &  - \frac{\lm^2g_A^2}{4\F} 
[(\ps+m)\gm_\al (\ps+\ks-m)\ks\pr (\ps\pr+m)]_{AB} 
\nonumber \\
& & \frac{i^3}{(p^2-m^2_N)(p'^2-m^2_N)\{(p+k)^2-m^2_N \}}
\chi^{a \dag}\tau^m\tau^n\chi^b +\mathrm{crossed\,\, term}
\eea
We evaluate it in the forward
direction for $\vec{p} =0$ and set $p_0\equiv E$ to get
\bea 
\frac{1}{\F}k^\al S^{ab}_\al & = &-6\de^{ab}\left( \frac{\lm g_A}{2\F}\right)^2 
p\d k(\ps+m)\ks(\ps+m)\frac{i^3}{(p^2-m^2_N)^2} 
\left( \frac{1}{p^2-m^2_N+2p\d k} + \frac{1}{p^2-m^2_N-2p\d k}\right) \\
& = & -6i\de ^{ab}\left(\frac{\lm^2 g_A^2}{2\F}\right)^2 
\frac{1}{E-m}\frac{1}{2} (1+\gm_0)\,,
\eea
in the vicinity of the pole. Inserting the results (3.5) and (3.9) into an
equation similar to Eq. (2.1),  the corrections change the free pole term
in vacuum
\[-\frac{\lm^2}{E-m+i\ep} \frac{1}{2}(1+\gm_o)\,, \]
to one with $\lm$ replaced by $\lm^T$ given by eq.(1.2).

\section{Discussion}

In this work we use the method of current algebra to derive the temperature
dependence of certain couplings, that have already been worked out with chiral
perturbation theory \cite{Mallik,Leutwyler1}. Basic to the former approach is
the extraction of terms of the appropriate amplitudes containing
the axial-vector current operator, that are singular as the momentum carried 
by this operator goes to zero. Generally speaking, the use of PCAC can 
incorporate in the calculation only a part of such singular pieces, namely 
those due to the pion pole. But there can be other singular pieces in the 
amplitude from heavy particle poles also, that must be calculated separately. 
The calculation of of the authors of Ref.\cite{Ioffe1} is incomplete in that 
it does not take the latter contribution into account, explaining the 
discrepancy between their results based apparently on current algebra and PCAC, 
and that of the effective theory. 

In the case of the two-point function of vector currents, the contribution
from the $\om$ meson pole can be as dominant as the pion pole in the 
neighbourhood of the $\rho$ meson pole. To be specific, it brings in
a small scale $\Delta m $ in the expression. For temperatures small compared 
to $\Delta m$, the $\om$ meson contribution is $O(T^4)$, changing the mass and 
residue only to this order. As $T$ increases past a narrow transition region 
around  $T =\Delta m$, the behaviour changes over to $O(T^2)$, when $\Fr^T$ is 
given by Eq. (1.1).  A similar result holds also for the axial-vector meson $a_1$ 
\cite{Mallik}. For the two point function of nucleon currents, it is the nucleon 
pole that must be included. Here the situation is simpler, as there is no mass 
difference in the problem. Accordingly $\lm^T$ is given by Eq.(1.2) for all 
temperatures.

It is interesting to compare here this old method of current algebra to
its modern development, namely that of effective field theory. The Ward 
identities used in the former method relate ampltudes with different number 
of pions, while in the effective theory one calculates the amplitudes directly 
in perturbation expansion.  
When one is in need of vertices involving heavy particles in 
a current algebra calculation, he has, however, to turn to the effective 
chiral Lagrangian. The insertion of intermediate states is rather clumsy 
when compared with the use of propagators in the effective field theory.
Also the two terms in Eqs. (2.13) and (3.8) can be obtained neatly 
from the mass shell delta function in the temperature dependent part of the 
pion propagator \cite{Mallik,Leutwyler1}. Thus even at the tree level, at
which current algebra works, one cannot but appreciate the technical and 
conceptual superiority of the method of effective field theory.

\section*{Acknowledgement}
The author is grateful to Prof. H. Leutwyler for suggesting the method 
of calculation presented here. He wishes to thank the members of 
the Institute for Theoretical Physics at the University of Berne for their
warm hospitality. He also acknowledges support of CSIR, Government of India.

\end{document}